\documentclass[11pt]{revtex4}

\usepackage{hyperref}
\usepackage{amsmath,amsfonts,amssymb}
\usepackage{braket}
\hypersetup{dvips,dvipdfm,colorlinks=true,urlcolor=magenta,filecolor=magenta,linktoc=page,citecolor=red,linkcolor=blue,bookmarks=true}
\usepackage{graphicx,epstopdf}

\begin{document}
	
	\title{ Noether Symmetry analysis in Chameleon Field Cosmology}
	\author{Roshni Bhaumik$^1$\footnote {roshnibhaumik1995@gmail.com}}
	\author{Sourav Dutta$^2$\footnote {sduttaju@gmail.com}}
	\author{Subenoy Chakraborty$^1$\footnote {schakraborty.math@gmail.com}}
	\affiliation{$^1$Department of Mathematics, Jadavpur University, Kolkata-700032, West Bengal, India\\$^2$Department of Mathematics, Dr. Meghnad Saha College, Itahar, Uttar Dinajpur-733128, West Bengal, India.}

	
	\begin{abstract}
		This work deals with chameleon field cosmology (a scalar field nonminimally coupled to cold dark matter) in the background of flat Friedmann-Lemaitre-Robertson-Walker (FLRW) space-time. Both classical and quantum cosmology have been investigated using Noether symmetry analysis of the underlying physical system. The Wheeler-DeWitt (WD) equation has been constructed on the minisuperspace and solutions have been obtained using conserved charge.\\	
	\end{abstract}
	\maketitle
	\textbf{Keywords}: Noether Symmetry; quantum cosmology; chameleon scalar field.

\section{Introduction}
Standard cosmology has been facing a great challenge since the end of the last century. The observational evidences since 1998 \cite{r1, r2, r3, r4, r5} are not in favour of decelerated expansion (prediction of standard cosmology) rather they are in favour of accelerated expansion. So far there are two options proposed by the cosmologists to accomodate these observational evidences. One of the possibilities is to introduce some exotic matter (known as dark energy (DE)) within the framework of Einstein gravity. This mysterious matter component is totally of unknown nature except its large negative pressure. At first cosmologists took cosmological constant as the DE candidate. But due to two severe drawbacks (namely discrepancy in its predicted and observed value and coincidence problem \cite{r6}) the cosmological constant is not a well accepted DE model rather dynamical DE models \cite{r7, r8, r9, r10} are widely used in the literature. This work is an example of using dynamical DE model. Usually, a scalar field having potential $V(\phi)$ is chosen as DE candidate so that the pressure component $p_{\phi}=\frac{1}{2}\dot{\phi}^2-V(\phi)$ can evolve to have the required negative value for observed accelerated expansion. Here, the scalar field (chosen as dynamical DE) is nonminimal coupled to dark matter (DM) through an interference term in the action \cite{r11}. As a result, there is a new source term in the matter conservation equation. This kind of DE model is termed as chameleon field. This model is quite useful to obtain accelerated expansion of the universe and other interesting cosmological consequences \cite{r12} (for details see the Ref. \cite{r13}).

On the other hand, since the last century, symmetry analysis has a significant role in studying global continuous symmetries (i.e, translation and rotation) as well as in local gauge symmetries, internal symmetries of the space-time (in cosmology) and permutation symmetry in quantum field theory \cite{r14, r15}. In particular, geometrical symmetries of the space-time and symmetries of the physical system have great role in analyzing any physical motion. From the perspective of Noether symmetry, the conserved charge has been used to identify the actual one among similar physical processes. Furthermore, in Noether symmetry approach, the Noether integral (i.e, the first integral) has been chosen as a tool for simplification of a system of differential equations or for the integrability of the system \cite{r16, r17, r18, r19, r20, r21}.

In addition, an advantage of using Noether symmetry to any physical system involving arbitrary physical parameters or some arbitrary functions of the field variables is that symmetry analysis uniquely determines these physical parameters or arbitrary functions involved (for details see Ref.\cite{r22}). Also since recent past symmetry analysis has been used for physical systems in Riemannian spaces \cite{r23, r24, r25, r26, r27, r28, r29, r1.3}. 

Moreover, Noether symmetry analysis has opened new windows in studying quantum cosmology with suitable operator ordering, and the Wheeler DeWitt (WD) equation so constructed on the minisuperspace is associated with Lie point symmetries. It is possible to have a subset of the general solutions of the WD equation having oscillatory behaviours \cite{r30, r31} by imposing Noether symmetries. The Noether symmetries with Hartle criterion can identify those classical trajectories in minisuperspace \cite{r32, r33} which are solutions of the cosmological evolution equations i.e, one may consider Noether symmetries as a bridge between quantum cosmology and classical observable universe.

This work is another example of extensive use of Noether symmetry analysis to both classical and quantum cosmology for chameleon field DE model. By imposing Noether symmetry to the Lagrangian and making canonical transformation of the dynamical variables it is possible to have classical solutions of the coupled nonlinear Einstein field equations. WD equation is constructed for the present chameleon DE cosmological model in the background of FLRW space-time and Noether symmetry is used as a tool to solve the WD equation. The plan of the paper is as follows: a brief overview of Noether symmetry is described in Section--II whereas Section-- III presents the Noether symmetry and cosmological solutions to chameleon field DE model and Section-IV deals with quantum cosmology in the minisuperspace approach: a general prescription and the formation of WD equation in the present cosmological model and possible solution with Noether symmetry are presented in Section-V; finally the paper ends with a conclusion in Section-VI.

\section{A Brief Overview of Noether Symmetry Approach}
Noether's first theorem states that any physical system is associated with some conserved quantities provided the Lagrangian of the system is invariant with respect to the Lie derivative \cite{r34, r35} along an appropriate vector field ($\mathcal{L}_{\overrightarrow{V}}f=\overrightarrow{V}(f)$). By imposing these symmetry constraints, the evolution equations of the physical system can either be solvable or simplified to a great extent \cite{r36, r37}.

For a point like canonical Lagrangian $L[q^{\alpha}(x^i),\dot{q}^{\alpha}(x^i)]$, the Euler-Lagrange equations 
\begin{equation}
	\partial_{j}\left(\frac{\partial L}{\partial\partial_{j}q^{\alpha}}\right)=\frac{\partial L}{\partial q^{\alpha}}\label{eqn1}
\end{equation}
can be contracted with some unknown functions $\lambda^{\alpha}(q^{\beta})$ as follows:
\begin{equation}
	\lambda^{\alpha}\bigg[\partial_{j}\left(\frac{\partial L}{\partial\partial_{j}q^{\alpha}}\right)-\frac{\partial L}{\partial q^{\alpha}}\bigg]=0\label{eqn2}
\end{equation}
i.e,
\begin{equation}
	\lambda^{\alpha}\frac{\partial L}{\partial q^{\alpha}}+(\partial_{j}\lambda^{\alpha})\left(\frac{\partial L}{\partial\partial_{j}q^{\alpha}}\right)=\partial_{j}\left(\lambda^{\alpha}\frac{\partial L}{\partial\partial_{j}q^{\alpha}}\right)\nonumber
\end{equation}
Thus
\begin{equation}
	\mathcal{L}_{\overrightarrow{X}}L=\lambda^{\alpha}\frac{\partial L}{\partial q^{\alpha}}+(\partial_{j}\lambda^{\alpha})\frac{\partial L}{\partial\left(\partial_{j}q^{\alpha}\right)}=\partial_{j}\left(\lambda^{\alpha}\frac{\partial L}{\partial\partial_{j}q^{\alpha}}\right)\label{eqn3}
\end{equation}
So according to Noether theorem the vector field \cite{r38, r39} 
\begin{equation}
	\overrightarrow{X}=\lambda^{\alpha}\frac{\partial}{\partial q^{\alpha}}+\left(\partial_{j}q^{\alpha}\right)\frac{\partial}{\partial\left(\partial_{j}q^{\alpha}\right)}\label{eqn4}
\end{equation}
can be chosen appropriately so that the Lagrangian of the system is invariant along the vector field i.e, $\mathcal{L}_{\overrightarrow{X}}L=0$ and consequently, the physical system is called invariant under Noether symmetry with $\overrightarrow{X}$, the infinitesimal generator of the symmetry. It is to be noted that the above symmetry vector as well as the Lagrangian is defined on the tangent space of configurations: $TQ\{q^{\alpha}, \dot{q}^{\alpha}\}$. In general Noether symmetry approach is very much relevant to identify conserved quantities of a physical system. The above symmetry condition is associated with a constant of motion for the Lagrangian having conserved phase flux along the vector field $\overrightarrow{X}$. Furthermore, from Eq.(\ref{eqn3}) this symmetry criteria is associated with a constant of motion of the system \cite{r16, r17, r36}
\begin{equation}
	Q^i=\lambda^{\alpha}\frac{\partial L}{\partial\left(\partial_{i}q^{\alpha}\right)}\label{eqn5}
\end{equation}
satisfying
\begin{equation}
	\partial_{i}Q^i=0\label{eqn6}
\end{equation}
So $Q^i$ is identified as Noether current or conserved current. Furthermore, the energy function associated with system is 
\begin{equation}
	E=\dot{q}^{\alpha}\frac{\partial L}{\partial\dot{q}^{\alpha}}-L\label{eqn7}
\end{equation}
The energy function (also known as Hamiltonian of the system) is a constant of motion provided there is no explicit time dependence in the Lagrangian \cite{r16, r17, r36}. Moreover, if the conserved current due to Noether symmetry has some physical meaning, \cite{r16, r17, r36} then symmetry analysis can identify reliable models. In the following, we shall show how symmetry analysis will simplify the present coupled cosmological model and as a result classical cosmological solutions can be obtained easily.

In the context of quantum cosmology, Hamiltonian formulation is very useful and Noether symmetry condition is rewritten as follows: \cite{r37}
\begin{equation}
	\mathcal{L}_{\overrightarrow{X}_{H}}H=0\label{eqn8}
\end{equation}
with
\begin{equation}
	{\overrightarrow{X}_{H}}=\dot{q}\frac{\partial}{\partial q}+\ddot{q}\frac{\partial}{\partial\dot{q}}\nonumber
\end{equation}
In minisuperspace models of quantum cosmology, symmetry analysis determines appropriate interpretation of the wave function. The conserved canonically conjugate momenta due to Noether symmetry can be written as follows:
\begin{equation}
	\Pi_{l}=\frac{\partial L}{\partial q^l}={\sum}_{l}\label{eqn9}
\end{equation}
$l=1,2,...,m,$
where `$m$' denotes the number of symmetries. Also, the operator version (i.e, quantization) of Eq. (\ref{eqn9}) i.e,
\begin{equation}
	-i\partial_{q^l}\ket{\psi}={\sum}_{l}\ket{\psi}\label{eqn10}
\end{equation}
identifies a translation along $q^l$-axis through symmetry analysis. Also Eq. (\ref{eqn10}) has oscillatory solution for real conserved quantity $\sum_{l}$ i.e,
\begin{equation}
	\ket{\psi}=\sum_{l=1}^{m}e^{i\sum_{l}q^l}\ket{\phi(q^k)}  , k<n,\label{eqn11}
\end{equation}
where the index `$k$' stands for directions along which there is no symmetry with $n$ the dimension of the minisuperspace. Thus oscillatory part of the wave function implies existence of Noether symmetry and the conjugate momenta along the symmetry directions should be conserved and vice-versa \cite{r41}. Due to symmetries the first integrals of motion identify the classical trajectories. In fact, for 2D minisuperspace, it is possible to have complete solution of the system by Noether symmetry.

\section{Noether Symmetry and Cosmological Solutions to Chameleon Field DE Model}
This section is devoted to study chameleon field DE cosmological model. This model consists of a canonical scalar field (having self-interaction potential) nonminimally coupled to DM. So the potential function and the coupling function are the two unknown functions of the scalar field. The action integral of the model has the explicit form \cite{r42, r43}
\begin{equation}
	I=\int\left[\frac{R}{16\pi G}+\frac{1}{2}\phi_{,\mu}\phi^{,\mu}-V(\phi)+f(\phi)L_{m}\right]\sqrt{-g}d^4x\label{eq1}
\end{equation}
where as usual $R$ is the Ricci scalar, $G$ is the Newtonian gravitational constant and $\phi$ is the chameleon scalar field having potential $V(\phi)$. Here, $L_{m}$ is the Lagrangian for DM which is nonminimally coupled to the chameleon scalar field with $f(\phi)$ (an analytic function), the coupling function. By choosing the DM to be an ideal gas, the matter Lagrangian can be chosen as $L_m\simeq\rho_{m}$ \cite{r1.1}.

In the background of flat FLRW space-time the point-like Lagrangian for the above cosmological model takes the following form:
\begin{equation}	
		L(a,\dot{a},\phi,\dot{\phi})=3a\dot{a}^2-a^3\left(\frac{\dot{\phi}^2}{2}-V(\phi)\right)-\rho_{m} f(\phi)a^{3}\label{eq2}
\end{equation}
Now the Euler-Lagrange equations (i.e, the Einstein field equations) for the Lagrangian (\ref{eq2}) are given by
\begin{eqnarray}
	3H^2=\rho_{m}f(\phi)+\frac{1}{2}\dot{\phi}^2+V(\phi)\label{eq3},\\
	{2\dot{H}+3H^2=-\frac{1}{2}\dot{\phi}^2+V(\phi)+\rho_{m}\omega f(\phi)},\label{eq4}
\end{eqnarray} 
where an over dot indicates differentiation with respect to the cosmic time `$t$'. {Furthermore, the equation of motion $T^{\mu\nu}_{;\nu}=0$ for the cosmological fluid with energy momentum tensor $T_{\mu\nu}=T^{(\phi)}_{\mu\nu}+T^{(m)}_{\mu\nu}$ is given by}
\begin{equation}
	{	\dot{\phi}\ddot{\phi}+3H\dot{\phi}^2+v(\phi)\dot{\phi}+\rho_{m}f'(\phi)\dot{\phi}+\dot{\rho_{m}}f(\phi)+3H(1+\omega)\rho_{m}f(\phi)=0}\label{eq5}
\end{equation}
One may note that among these three evolution equations (\ref{eq3})-(\ref{eq5}), only two are independent while (\ref{eq3}) is termed as constraint equation.

{As in the present cosmological model there is the interaction term $f(\phi)$ so one has $\left(T^{(\phi)\mu\nu}\right)_{;\nu}=-Q$, $\left(T^{(m)\mu\nu}\right)_{;\nu}=Q$ or eqivalently }
\begin{equation}
	{\ddot{\phi}+3H\dot{\phi}+V'(\phi)+f'(\phi)\rho_{m}=-\frac{Q}{\dot{\phi}}}\label{eq17.1}
\end{equation}
{and}
\begin{equation}
	{	\dot{\rho_{m}}f(\phi)+3H(1+\omega)\rho_{m}f(\phi)=Q}\label{eq18.1}
\end{equation}
{As the present model reduces to that of Weyl integrable gravity with $f(\phi)=f_0e^{\lambda\phi}$ (\cite{r1.2}) and setting $Q=\alpha\rho_{m}f'(\phi)\dot{\phi}$ (where $\alpha$ is a non zero constant), Eqs. (\ref{eq17.1}) and (\ref{eq18.1}) become}
\begin{equation}
	{\ddot{\phi}+3H\dot{\phi}+V'(\phi)+(1+\alpha)f'(\phi)\rho_{m}=0}\label{eq17.2}
\end{equation}
and
\begin{equation}
	{	\dot{\rho_{m}}f(\phi)+3H(1+\omega)\rho_{m}f(\phi)-\alpha f'(\phi)\dot{\phi}\rho_{m}=0}\label{eq18.2}
\end{equation}
{The matter conservation equation (\ref{eq18.2}) can be integrated to have }
\begin{equation}
	{\rho_{m}(t)=\rho_{0}a^{-3(1+\omega)}\{f(\phi)\}^{\alpha}}\label{eq21.1}
\end{equation}
{Thus the scalar field evolution equation (i.e, the modified Klein-Gordon equation) (\ref{eq17.2}) becomes}
\begin{equation}
	{	\ddot{\phi}+3H\dot{\phi}+V'(\phi)=-\rho_{0}a^{-3(1+\omega)}\left[\{f(\phi)\}^{\alpha+1}\right]}\label{eq22.1}.
\end{equation}

The configuration space for the present model is a 2D space $(a,\phi)$ and the infinitesimal generator for the Noether symmetry takes the form 
\begin{equation}
	\overrightarrow{X}=p\frac{\partial}{\partial a}+q\frac{\partial}{\partial \phi}+\dot{p}\frac{\partial}{\partial \dot{a}}+\dot{q}\frac{\partial}{\partial\dot{\phi}},\label{eq6}
\end{equation}
where $p=p(a,\phi)$ and $q=q(a,\phi)$ are the unknown coefficients with
$\dot{p}=\frac{\partial p}{\partial a}\dot{a}+\frac{\partial p}{\partial\phi}\dot{\phi}$ and similarly for $\dot{q}$.

These coefficients of the Noether symmetry vector are determined from an overdetermined system of partial differential equation, obtained by imposing Noether symmetry to the Lagrangian i.e,
\begin{equation}
	\mathcal{L}_{\overrightarrow{X}}L=0\nonumber
\end{equation}
i.e,
\begin{eqnarray}
	p+2a\frac{\partial p}{\partial a}&=&0\nonumber\\
	3p+2a\frac{\partial q}{\partial\phi}&=&0\nonumber\\
	6\frac{\partial p}{\partial\phi}-a^2\frac{\partial q}{\partial a}&=&0\label{eq7}
\end{eqnarray}
with a differential equation for the potential and coupling function as follows:
\begin{equation}
	{3\rho_{0}\omega pa^{-3\omega-1}F(\phi)+3pa^2V(\phi)+qa^3 V'(\phi)-\rho_{0} a^{-3\omega}qF'(\phi)=0}\label{eq8}
\end{equation}
{where ${F(\phi)=\{f(\phi)\}^{\alpha+1}}$}.

The above set of partial differential equations (\ref{eq7}) are solvable using the method of separation of variables i.e, $p(a,\phi)=p_{1}(a)p_{2}(\phi)$, $q(a,\phi)=q_{1}(a)q_{2}(\phi)$ as
\begin{equation}
	p=a^{-\frac{1}{2}}\left(c_{p}e^{m\phi}+c_{q}e^{-m\phi}\right)\nonumber
\end{equation}
\begin{equation}	
	{q=-4ma^{-\frac{3}{2}}\left(c_{p}e^{m\phi}-c_{q}e^{-m\phi}\right)}\label{eq9}
\end{equation}
where $m^2=\frac{3}{8}$, $c_{p}$, $c_{q}$ and $q_{0}$ are arbitrary constants. Using the above solutions (\ref{eq9}) into (\ref{eq8}), the solutions for $V(\phi)$ and $f(\phi)$ can take the form (with $\omega=-1$) 
\begin{equation}
	{ V(\phi)-\rho_{0}F(\phi)=k\left(c_{p}e^{m\phi}-c_{q}e^{-m\phi}\right)^2}\label{eq27.1}
\end{equation}
{where $k$ is a positive integration constant.}

Thus, the infinitesimal generator of the Noether symmetry is determined (except for arbitrary integration constants) by imposing symmetry condition which in turn determines a relation between the potential function and the coupling function.

Another important issue related to Noether symmetry is the conserved quantities associated with it. In general for a field theory in curved space there is no well-defined notion of energy. However, the conserved quantity derived from Noether's theorem is the energy-momentum tensor. In particular, when the system has time-like killing vector then there is an associated conserved energy. Though FLRW space-time has no time-like killing vector field, but the Lagrangian density is explicit time independent. Hence in analogy with point-like Lagrangian, it is possible to define an energy which will be conserved in nature. Thus in the context of Noether symmetry to the present cosmological model one can have two conserved quantities, namely conserved charge (defined in Eq. (\ref{eqn5})) and conserved energy (defined in Eq. (\ref{eqn7})) having explicit form 
\begin{equation}
	{Q=6\dot{a}a^{\frac{1}{2}}\left(c_{p}e^{m\phi}+c_{q}e^{-m\phi}\right)+a^3\dot{\phi}\bigg\{4ma^{-\frac{3}{2}}\left(c_{p}e^{m\phi}-c_{q}e^{-m\phi}\right)\bigg\}}\nonumber
\end{equation}
\begin{equation}
	{E=3a\dot{a}^2-\frac{1}{2}a^3\dot{\phi}^2-a^3V(\phi)+\rho_{0}F(\phi)a^{-3\omega}}\label{eq12}
\end{equation}
Usually, associated with Noether symmetry there is a conserved current (defined in Eq. (\ref{eqn5})), whose time component integrating over spatial volume gives a conserved charge. But in the present context as all the variables are time dependent only so $Q$ defined in (\ref{eq12}) is the Noether charge. Moreover, the above conserved charge can be expressed geometrically as the inner product of the infinitesimal generator with cartan one form \cite{r44} as follows:
\begin{equation}
	Q=i_{\overrightarrow{X}}\theta_{L}\label{eq13}
\end{equation} 
where $i_{\overrightarrow{X}}$ denotes the inner product with the vector field $\overrightarrow{X}$ and
\begin{equation}
	\theta_{L}=\frac{\partial L}{\partial a}da+\frac{\partial L}{\partial\phi}d\phi\label{eq14}
\end{equation}
is termed as cartan one form.

On the other hand, this geometric inner product representation is useful to find out cyclic variables in the Lagrangian. In context of solving coupled nonlinear evolution equations, determination of cyclic variables will be very useful as not only the Lagrangian but also the evolution equations will be simplified to a great extent.

In the present context the transformation of the 2D augmented space: $(a,\phi)\rightarrow(u,v)$ transform the symmetry vector as
\begin{equation}
	\overrightarrow{X_{T}}=\left(i_{\overrightarrow{X}}du\right)\frac{\partial}{\partial u}+\left(i_{\overrightarrow{X}}dv\right)\frac{\partial}{\partial v}+\left\{\frac{d}{dt}\left(i_{\overrightarrow{X}}du\right)\right\}\frac{d}{d\dot{u}}+\left\{\frac{d}{dt}\left(i_{\overrightarrow{X}}dv\right)\right\}\frac{d}{d\dot{v}}\label{eq15}
\end{equation}
Geometrically, $\overrightarrow{X_{T}}$ may be interpreted as the lift of a vector field on the augmented space. Now, without any loss of generality we restrict the above point transformation to \cite{r44}
\begin{equation}
	i_{\overrightarrow{X}}du=1~~\mbox{and}~~	i_{\overrightarrow{X}}dv=0\label{eq16}
\end{equation}
so that
\begin{equation}
	\overrightarrow{X_{T}}=\frac{\partial}{\partial u}~~\mbox{and}~~\frac{\partial L_{T}}{\partial u}=0\label{eq17}
\end{equation}

i.e, $u$ is the cyclic variable. The above geometric process of identification of cyclic variables can be interpreted so as to choose the transformed infinitesimal generator along any co-ordinate line (identified as the cyclic variable) \cite{r45}.

Now the explicit form of the above point transformation (\ref{eq16}) is the first-order linear partial differential equations having solution as follows:\\

\textbf{\underline{Case I:}} $c_{p}=c_{q}$
\begin{eqnarray}
	u&=&\frac{2}{3}a^{\frac{3}{2}}\cosh{m\phi},\nonumber\\
	v&=&a^{\frac{3}{2}}\sinh{m\phi}\label{eq18}.
\end{eqnarray}
\textbf{\underline{Case II:}} $c_{p}\neq c_{q}$
\begin{eqnarray}
	u&=&\frac{1}{6c_{p}c_{q}}a^{\frac{3}{2}}\left(c_{p}e^{m\phi}+c_{q}e^{-m\phi}\right),\nonumber\\
	v&=&a^{\frac{3}{2}}\left(c_{p}e^{m\phi}-c_{q}e^{-m\phi}\right)\label{eq19}.
\end{eqnarray}
The simplified Lagrangian in the new variables has the following forms:
\begin{eqnarray}
	L&=&3\dot{u}^2-\frac{4}{3}\dot{v}^2+4kc_{p}^2v^2\label{l1},~~~~~ (\mbox{Case I})\\
	&=&12c_{p}c_{q}\dot{u}^2-\frac{1}{3c_{p}c_{q}}\dot{v}^2+kv^2\label{l2}  ~~~~~ (\mbox{Case II}).
\end{eqnarray}
The conserved quantities in the new variables can be expressed as follows:
\begin{eqnarray}
	Q&=&6\dot{u}\nonumber\\
	E&=&3\dot{u}^2-\frac{4}{3}\dot{v}^2-4kc_{p}^2v^2\nonumber  ~~~~~(\mbox{Case I})
\end{eqnarray}
and
\begin{eqnarray}
	Q&=&24c_{p}c_{q}\dot{u}\nonumber\\
	E&=&12c_{p}c_{q}\dot{u}^2-\frac{1}{3c_{p}c_{q}}\dot{v}^2-kv^2\nonumber  ~~~~~(\mbox{Case II})
\end{eqnarray}
Now solving the Euler-Lagrange equations for the new Lagrangian, the new augmented variables have the following forms:
\begin{eqnarray}
	u&=&At+B\nonumber\\
	v&=&k_{1}\cos\sqrt{3k}c_{p}t+k_{2}\sin\sqrt{3k}c_{p}t\nonumber   ~~~~~~ (\mbox{Case I})
\end{eqnarray}
and
\begin{eqnarray}
	u&=&rt+s\nonumber\\
	v&=&k'_{1}\cos\sqrt{3c_{p}c_{q}k}t+k'_{2}\sin\sqrt{3c_{p}c_{q}k}\nonumber   ~~~~~(\mbox{Case II})
\end{eqnarray}
Hence, the cosmic scale factor and the chameleon scalar field have the following explicit expressions:
\begin{eqnarray}
	a(t)&=&\left[\frac{9}{4}\left(At+B\right)^2-\left(k_{1}\cos\sqrt{3k}c_{p}t+k_{2}\sin\sqrt{3k}c_{p}t\right)^2\right]^{\frac{1}{3}}\nonumber\\
	\phi(t)&=&\frac{2\sqrt{2}}{3}\tanh^{-1}\left[\frac{2\left(k_{1}\cos\sqrt{3k}c_{p}t+k_{2}\sin\sqrt{3k}c_{p}t\right)}{3(At+B)}\right]\nonumber    ~~~~~(\mbox{Case I})
\end{eqnarray}
and
\begin{eqnarray}
	a(t)&=&\left[9c_{p}c_{q}\left(rt+s\right)^2-\frac{1}{4c_{p}c_{q}}\left(k'_{1}\cos\sqrt{3c_{p}c_{q}k}t+k'_{2}\sin\sqrt{3c_{p}c_{q}k}t\right)^2\right]^\frac{1}{3}\nonumber\\
	\phi(t)&=&\frac{2\sqrt{2}}{\sqrt{3}}\ln\frac{6c_{p}c_{q}(rt+s)+\left(k'_{1}\cos\sqrt{3c_{p}c_{q}k}t+k'_{2}\sin\sqrt{3c_{p}c_{q}k}t\right)}{2c_{p}\left[9c_{p}c_{q}\left(rt+s\right)^2-\frac{1}{4c_{p}c_{q}}\left(k'_{1}\cos\sqrt{3c_{p}c_{q}k}t+k'_{2}\sin\sqrt{3c_{p}c_{q}k}t\right)^2\right]^\frac{1}{2}}\nonumber    ~~~~~(\mbox{Case II})
\end{eqnarray}

In the above solutions $(A, B, k_{1}, k_{2})$ and $(r, s, k'_{1}, k'_{2})$ are arbitrary integration constants.
\begin{figure}
	\centering
	\includegraphics[width=0.5\textwidth]{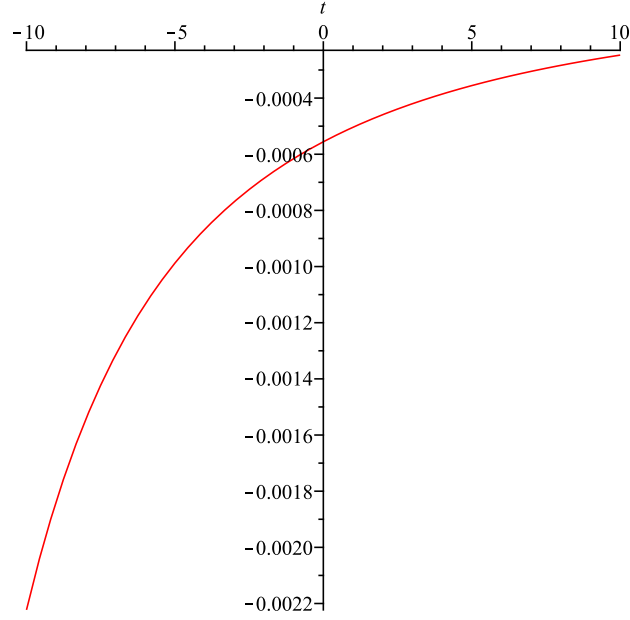}\\
	\caption{Graphical representation of $\frac{\ddot{a}}{a}$ with respect to cosmic time $t$ when $c_{p} = c_{q}$.}
	\label{fig1}
\end{figure}
\begin{figure}
	\centering
	\includegraphics[width=0.5\textwidth]{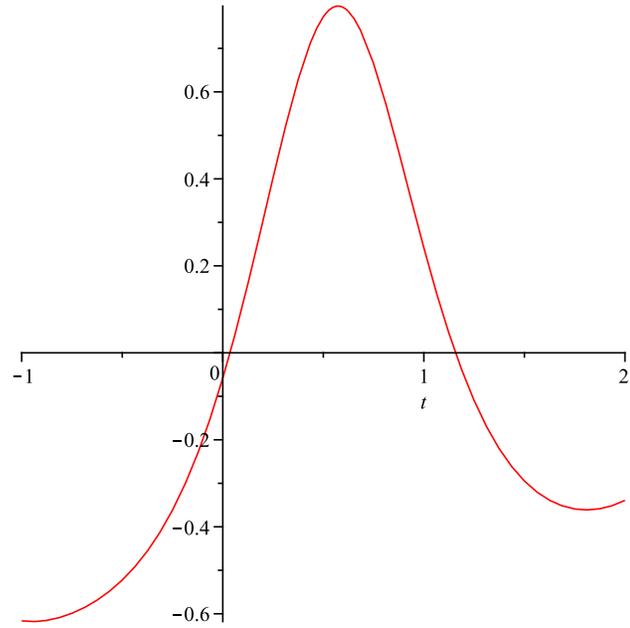}\\
	\caption{Represents $\frac{\ddot{a}}{a}$ with respect to cosmic time $t$ when $c_{p} \neq c_{q}$.}
	\label{fig2}
\end{figure}\\

\section{Quantum Cosmology in the Minisuperspace Approach: A General Prescription}
Minisuperspaces are considered as restrictions of geometrodynamics of the superspace and physically important and interesting models are defined on minisuperspaces. In cosmology, the simplest and widely used minisuperspace models are homogeneous and isotropic merics and matter fields and consequently the lapse function is homogeneous (i.e, $N=N(t)$) while shift function vanishes identically. So in 4D manifold, using $(3+1)$-decomposition the metric can be written as follows:
\begin{equation}
	ds^2=-N^2(t)dt^2+h_{ab}(x,t)dx^adx^b\label{eqq1}
\end{equation}   
and the Einstein-Hilbert action can be written as follows:
\begin{equation}
	I(h_{ab},N)=\frac{{m^2}_p}{16\pi}\int dt~d^3xN\sqrt{h}\left[k_{ab}k^{ab}-k^2+(3)_{R}-2\Lambda\right]\label{eqq2},
\end{equation}
where $k_{ab}$ is the extrinsic curvature of the $3$ space, $k=k_{ab}h^{ab}$ is the trace of the extrinsic curvature, $(3)_{R}$ is the curvature scalar of the three space and $\Lambda$ is the cosmological constant.

Now due to homogeneity of the three space, the metric $h_{ab}$ is characterized by a finite number of time functions $q^{\alpha}(t)$, $\alpha=0,1,2,...,n-1$ and the above action can be written in the form of a relativistic point particle with self-interacting potential in a $n$D curved space-time as \cite{r42, r43}
\begin{equation}
	I\left(q^{\alpha}(t),N(t)\right)=\int_{0}^{1}dtN\left[\frac{1}{2N^2}f_{\alpha\beta}(q)\dot{q}^{\alpha}\dot{q}^{\beta}-V(q)\right]\label{eqq3}
\end{equation} 
So the equation of motion of the (equivalent) relativistic particle can be written as (considering variation of the action with respect to the field variables $q^{\alpha}(t)$)
\begin{equation}
	\frac{1}{N}\frac{d}{dt}\left(\frac{\dot{q}^{\alpha}}{N}\right)+\frac{1}{N^2}\Gamma^{\alpha}_{\mu\nu}\dot{q}^{\mu}\dot{q}^{\nu}+f^{\alpha\beta}\frac{\partial\nu}{\partial q^{\beta}}=0\label{eqq4}
\end{equation}
with $\Gamma^{\alpha}_{\beta\gamma}$ being the christoffel symbols in the minisuperspace. Also there is a constraint equation obtained by variation with respect to the lapse function as follows:
\begin{equation}
	\frac{1}{2N^2}f_{\alpha\beta}\dot{q}^{\alpha}\dot{q}^{\beta}+V(q)=0\label{eqq5}
\end{equation}
For canonical quantization scheme one has to switch over to Hamiltonian formulation. The momenta canonical to $q^{\alpha}$ are given by 
\begin{equation}
	p_{\alpha}=\frac{\partial L}{\partial q^{\alpha}}=f_{\alpha\beta}\frac{\dot{q}^{\beta}}{N},\label{eqq6}
\end{equation}
so the Hamiltonian is defined as follows:
\begin{equation}
	H=p_{\alpha}\dot{q}^{\alpha}-L=N\left[\frac{1}{2}f^{\alpha\beta}p_{\alpha}p_{\beta}+V(q)\right]=N\mathcal{H}\label{eqq7}
\end{equation}
with $f^{\alpha\beta}$ being the inverse metric. Using the definition of $p_{\alpha}$ (i.e, equation (\ref{eqq6})) into the constraint equation (\ref{eqq5}) one obtains 
\begin{equation}
	\mathcal{H}(q^{\alpha},p_{\alpha})\equiv\frac{1}{2}f^{\alpha\beta}p_{\alpha}p_{\beta}+V(q)=0\label{eqq8}
\end{equation}
Now, writing $p_{\alpha}$ as $-i\hbar\dfrac{\partial}{\partial q_{\alpha}}$ in quantization scheme, the operator version of the above constraint equation (\ref{eqq8}) on a time-independent function (the wave function of the universe), one gets the WD equation in quantum cosmology as follows:
\begin{equation}
	\mathcal{H}\left(q^{\alpha},-i\hbar\frac{\partial}{\partial q^{\alpha}}\right)\psi(q^{\alpha})=0\label{eqq9}
\end{equation}
In general, the minisuperspace metric depends on $q^{\alpha}$, so the above WD equation has operator ordering problem. However by imposing the quantization in minisuperspace to be covariant in nature, one may resolve the above operator ordering problem. Furthermore, in the context of quantam cosmology for probability measure, $\exists$ a conserved current for hyperbolic type of partial differential equation is as follows: 
\begin{equation}
	\overrightarrow{J}=\frac{i}{2}(\psi^{*}\nabla\psi-\psi\nabla\psi^{*})\label{eqq10}
\end{equation}
with $\overrightarrow{\nabla}.\overrightarrow{J}=0$. Here, $\psi$ is the solution of the hyperbolic-type WD differential equation. Thus it is possible to define the probability measure on the minisuperspace as follows:
\begin{equation}
	dp=|\psi(q^{\alpha})|^2dV\label{eqq11}
\end{equation}
where $dV$ is a volume element on minisuperspace.

\section{Formation of WD Equation in the Present Cosmological Model and possible solution with Noether Symmetry}
In the present cosmological model, the 2D configuration space $\{a,\phi\}$ is associated with conjugate momenta is given by
\begin{eqnarray}
	p_{a}&=&\frac{\partial L}{\partial\dot{a}}=6a\dot{a}\nonumber\\
	p_{\phi}&=&\frac{\partial L}{\partial\dot{\phi}}=-a^3\dot{\phi}\label{eq20}
\end{eqnarray}
So the Hamiltonian of the system (also known as Hamiltonian constraint) can be expressed as follows:
\begin{equation}
	{\mathcal{H}=\frac{1}{12a}p_{a}^2-\frac{1}{2a^3}p_{\phi}^2-a^3V(\phi)+\rho_{0} F(\phi)a^{-3\omega}}\label{eq21}
\end{equation}   
with equivalent Hamilton's equations of motion
\begin{eqnarray}
	\dot{a}&=&\frac{1}{6a}p_{a}\nonumber\\
	\dot{\phi}&=&-\frac{1}{a^3}p_{\phi}\nonumber\\
	{\dot{p_{a}}}&=&{\frac{1}{12a^2}p_{a}^2-\frac{3}{2a^4}p_{\phi}^2+3a^2V(\phi)+3\rho_{0}\omega F(\phi)a^{-3\omega-1}}\nonumber\\
	{\dot{p_{\phi}}}&=&{a^3V'(\phi)-\rho_{0} F'(\phi)a^{-3\omega}}\label{eq22}
\end{eqnarray}
Furthermore, the Lagrangian (i.e, Eq. (\ref{eq2})) of the system can be interpreted geometrically, dividing it into two parts. The first two terms are known as kinetic part and the remaining two terms constitute the dynamic part. Also the kinetic part may be viewed as a 2D pseudo- Riemannian space with line element 
\begin{equation}
	ds^2=-6ada^2+a^2d\phi^2\label{eq23}
\end{equation}
This 2D Lorentzian manifold $(a,\phi)$ is known as minisuperspace (in quantum cosmology). The wave function of the universe in quantum cosmology is a solution of the WD equation, a second-order hyperbolic partial differential equation defined over minisuperspace and it is the operator version of the Hamiltonian constraint.

Furthermore, in the context of WKB approximation one can write the wave function as $\psi(x^k) \sim e^{i\delta(x^k)}$ and hence the WD equation (\ref{eqq9}) becomes first-order nonlinear partial differential equation which is nothing but (null) Hamilton-Jacobi (H-J) equation in the same geometry.

In quantization of the model one has to construct the WD equation $\hat{\mathcal{H}}\psi(u,v)=0$, with $\hat{\mathcal{H}}$ the operator version of the Hamiltonian  (\ref{eq21}) and $\psi(u,v)$, the wave function of the universe. In course of conversion to the operator version there is a problem related to the ordering of a variable and its conjugate momentum \cite{r46}. In the first term of the Hamiltonian  (\ref{eq21}) there is a product of `$a$' and `$p_a$', so one has to consider the ordering consideration: $p_a \rightarrow -i\partial_a$,~ $p_{\phi} \rightarrow -i\partial_{\phi}$. As a result there is a two-parameter family of WD equation 
\begin{equation}
	{\bigg[-\frac{1}{12} \frac{1}{a^l}\frac{\partial}{\partial a}\frac{1}{a^m}\frac{\partial}{\partial a}\frac{1}{a^n}+\frac{1}{a^3}\frac{\partial^2}{\partial \phi^2}-a^3 V(\phi)+\rho_0 F(\phi)a^{-3\omega}\bigg]\psi(a, \phi)=0}\label{eq24}
\end{equation}
with the triplet of real numbers $(l, m, n)$ satisfying $l+m+n=1$. Due to infinite possible choices for $(l, m, n)$ one may have infinite number of possible ordering. Also the semi-classical limit, namely, the Hamilton Jacobi equation (obtained by substituting $\psi=\exp (is))$ does not regard to the above triplet. In fact, the following choices are commonly used:\\

i) $l=2, m=-1, n=0$ : D'Alembert operator ordering.\\

ii) $l=0=n, m=1$: Vilenkin operator ordering.\\

iii) $l=1, m=0=n$: no ordering.\\

Thus factor ordering affects the behaviour of the wave function while semi classical results will not be influenced by the above ordering problem. Now choosing the third option (i.e., no ordering) the WD equation for the present model has the following explicit form:
\begin{equation}
	{\bigg[-\frac{1}{12a} \frac{\partial^2}{\partial a^2}+\frac{1}{2a^3}\frac{\partial^2}{\partial \phi^2}-a^3 V(\phi)+\rho_0 F(\phi)a^{-3\omega}\bigg]\psi(a, \phi)=0}\label{eq25}
\end{equation}

The general solution of the above second-order hyperbolic partial differential equation is known as the wave function of the universe. This solution can be constructed from the separation of the eigen functions of the above WD operator as follows:\cite{r41}
\begin{equation}
	\psi(a, \phi)=\int W(Q)\psi(a, \phi, Q)~dQ\label{eq26}
\end{equation}
with $\psi$ being an eigen function of the WD operator, $W(Q)$ being a weight function and $Q$ being the conserved charge. Now it is desirable to have wave function in quantum cosmology that is consistent with classical theory. In other words, one has to construct a coherent wave packet having good asymptotic behaviour in the minisuperspace and maximize around the classical trajectory. As the minisuperspace variables $\{a, \phi\}$ are highly coupled in the WD operator so it is not possible to have any explicit solution of the WD equation even with separation of variable method. Thus one may analyze the present model in the context of quantum cosmology using the new variables $(u, v)$ (obtained by point transformation) in the augmented space\\\\

{\bf{\underline{Case-I}}}: $c_p=c_q$\\

In this case the Lagrangian is given by Eq. (\ref{l1}) for which $u$ is the cyclic variable. So one has
\begin{eqnarray}
	p_1&=&\frac{\partial L}{\partial \dot{u}}=6\dot{u}=\mbox{Conserved}\nonumber\\
	p_2&=&\frac{\partial L}{\partial \dot{v}}=-\frac{8}{3}\dot{v}\label{eq27}
\end{eqnarray}
Hence, the Hamiltonian of the system takes the form
\begin{equation}
	\mathcal{H}=\frac{1}{12}p_{u}^2-\frac{3}{16}p_{v}^2-4kc_p^2v^2\label{eq28}
\end{equation} 
Thus the WD equation takes the following form:
\begin{equation}
	\bigg[-\frac{1}{12} \frac{\partial^2}{\partial u^2}+\frac{3}{16}\frac{\partial^2}{\partial v^2}-4kc_p^2v^2\bigg]\chi(u, v)=0\label{eq29}
\end{equation}
The operator version of the conserved momentum in Eq. (\ref{eq27}) can be written as follows:
\begin{equation}
	i \frac{\partial \chi(u, v)}{\partial u}=\Sigma_0 ~\chi(u, v)\label{eq30}
\end{equation}
Now writing $\chi(u, v)=A(u) B(v)$, one has
\begin{eqnarray}
	i\frac{dA}{du}&=&\Sigma_0~A\nonumber\\
	\mbox{i.e.,}~A(u)&=&A_0 \exp(-i\Sigma_0 ~u) \label{eq31}
\end{eqnarray}
with $A_0$ being the constant of integration. Using Eq. (\ref{eq31}) the WD equation (\ref{eq29})  becomes a differential equation in $B$ as follows:
\begin{eqnarray}
	\frac{3}{16} \frac{d^2B}{dv^2}-4kc_p^2v^2B+\frac{\Sigma_0^2}{12}B&=&0\nonumber\\
	\mbox{i.e.,}~\frac{d^2B}{dv^2}-(\lambda v^2-\mu)B&=&0\label{eq32}
\end{eqnarray}
with $\lambda=\frac{64}{3} kc_p^2,~\mu=\frac{4}{9}\Sigma_0^2$.\\

{\bf{\underline{Case-II}}}: $c_p \neq c_q$\\

The Lagrangian of the system (given by Eq. (\ref{l2})) shows the variable `$u$' to be cyclic and the conserved momentum has the following expression:
\begin{equation}
	p_u=\frac{\partial L}{\partial \dot{u}}=24 c_p c_q \dot{u}=\Lambda_0,~\mbox{a constant}
	\label{eq35}
\end{equation}
while the momentum conjugate to the variable `$v$' is given by
\begin{equation}
	p_2=\frac{\partial L}{\partial \dot{v}}=-\frac{2}{3c_p c_q}\dot{v}\label{eq36}
\end{equation}
Hence, the Hamiltonian of the system is expressed as follows:
\begin{equation}
	\mathcal{H}=\frac{1}{48c_p c_q}p_u^2-\frac{3c_pc_q}{4}p_v^2-kv^2\label{eq37}
\end{equation} 
and consequently the WD equation takes the following form:
\begin{equation}
	\bigg[\frac{1}{48c_p c_q}\frac{\partial^2}{\partial u^2}+\frac{3c_pc_q}{4}\frac{\partial^2}{\partial v^2}-kv^2\bigg]\xi(u, v)=0\label{eq38}
\end{equation}
The operator version of the conserved momentum as before shows
\begin{eqnarray}
	\xi(u, v)&=&C(u)D(v)\nonumber\\
	\mbox{with}~C(u)&=&C_0\exp (-i\Lambda_0 u)\label{eq39}
\end{eqnarray}
Thus from the above WD equation (\ref{eq38}) and using the separation of variables, the differential equation for $D$ reduces to 
\begin{equation}
	\frac{d^2D}{dv^2}-(l v^2-s)D=0\label{eq40}
\end{equation}
with $l=\frac{4k}{3c_p c_q} ,~s=\frac{\Lambda_0^2}{36 c_p^2 c_q^2}$.\\
The solution of this second-order differential equation takes the following form:
\begin{eqnarray}
	{D(v)}&=&{C_1 \sqrt{v}J_{\frac{1}{4}}\bigg( \frac{1}{2}\sqrt{-l}v^2\bigg)+C_2\sqrt{v}Y_{\frac{1}{4}}\bigg( \frac{1}{2}\sqrt{-l}v^2\bigg)~~~~\mbox{when}(s=0)}\nonumber\\
	&=&{\frac{C_1M_{\frac{1}{4}\frac{s}{\sqrt{l}},\frac{1}{4}}\bigg( \sqrt{l}v^2\bigg)}{\sqrt{v}}+\frac{C_2W_{\frac{1}{4}\frac{s}{\sqrt{l}},\frac{1}{4}}\bigg( \sqrt{l}v^2\bigg)}{\sqrt{v}}~~~~\mbox{when}(s\neq 0)}, \label{eq42}
\end{eqnarray}
{where $J$ and $Y$ are usual Bessel functions and $M$ and $W$ are known as Whittaker functions. We have represented the wave function graphically both for zero and non-zero $s$ in Figs.3 and 4, respectively. From  the figures,  we see that at $u=0$, $v=0$ (i.e, $l=0$), wave function has finite nonzero value. Therefore in the present model it is possible to avoid the Big Bang singularity using quantum cosmology near the initial singularity. }
\begin{figure}
	\centering
	\includegraphics[width=0.5\textwidth]{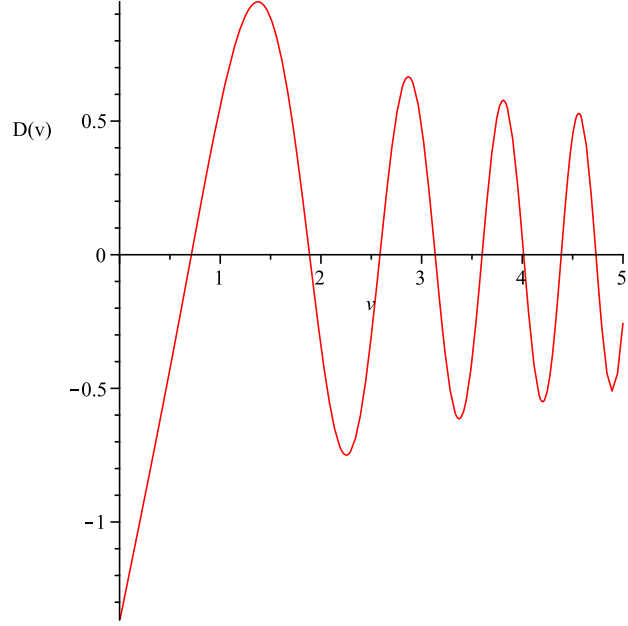}\\
	\caption {The wave function when $s=0$}
	\label{fig3}
\end{figure}
\begin{figure}
	\centering
	\includegraphics[width=0.5\textwidth]{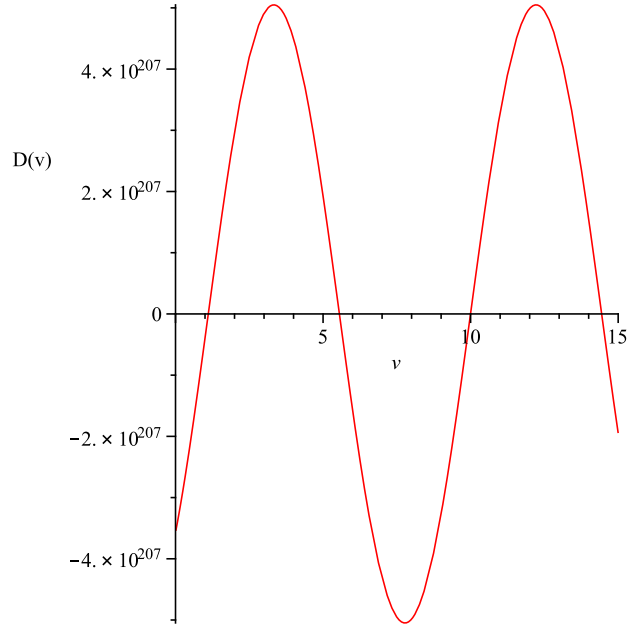}\\
	\caption {Graphical representation of the wave function when $s\neq 0$}
	\label{fig4}
\end{figure}

\section{Conclusion}
This work is an example where symmetry analysis particularly Noether symmetry has been extensively used both in classical and quantum cosmology. Here, chameleon field DE model has been considered in the background of homogeneous and isotropic flat FLRW space--time. Although the full quantum theory is described on the infinite-dimensional superspace, but here we shall confine to minisuperspace which is a 2D Lorentzian manifold.\\

Although the Einstein field equations are nonlinear coupled differential equation, but using a transformation in the augmented space and introducing geometric inner product it is possible to identify the cyclic variable(s) so that the field equations simplified to a great extent and consequently classical cosmological solutions are evaluated. There are two sets of solutions for two different choices of the arbitrary constants involved. Both the solutions show an expanding model of the universe with accelerating and decelerating phases (depending on the choices of the arbitrary constants involved). In particular, the present model describes the decelerating phase only for the choice $c_{p}=c_{q}$ (Fig. \ref{fig1}), while the model makes a transition from decelerating phase to accelerating phase and then again it goes to decelerating phase for the choice $c_{p} \neq c_{q}$ (Fig. \ref{fig2}).\\

On the other hand, the application of Noether symmetry to the minisuperspace shows the path for solving WD equation. The conserved momentum due to Noether symmetry, after converting to quantum version, shows an oscillatory solution to the WD equation and consequently it gives the semi-classical limit of quantum cosmology. Furthermore, the nonoscillatory part of the WD equation is an ordinary differential equation having solution in the form of Bessel function or Whittaker function. The graphical presentation of this part of the solution has been shown in Figs. (\ref{fig3} and\ref{fig4}), which clearly shows that the present quantum cosmological model can overcome the big bang singularity i.e., the present model may describe the early era of evolution without any singularity. Finally, one may conclude that Noether symmetry analysis is very useful in describing quantum cosmology in minisuperspace model and also leads to possible solution of the WD equation.

\end{document}